%% file: cas-sc-template.tex
\documentclass[a4paper,fleqn]{cas-sc}

\usepackage[numbers]{natbib}
\usepackage{lineno}
\usepackage{caption}
\usepackage{graphicx}
\usepackage{subfigure}
\usepackage{placeins}
\usepackage{multirow}
\usepackage{lineno}

\def\tsc#1{\csdef{#1}{\textsc{\lowercase{#1}}\xspace}}
\tsc{WGM}
\tsc{QE}
\tsc{EP}
\tsc{PMS}
\tsc{BEC}
\tsc{DE}

\begin{document}

\let\WriteBookmarks\relax
\def\floatpagepagefraction{1}
\def\textpagefraction{.001}

\shorttitle{Ambient Neutron Measurement at Taishan Antineutrino Observatory}

\shortauthors{Ruhui Li et~al.}

\title [mode = title]{Ambient Neutron Measurement at Taishan Antineutrino Observatory}


%
\author[1,2]{Ruhui Li}[type=editor,
                        auid=000,bioid=1,
                        prefix=,
                        role=,
                        orcid=]
\fnmark[1]

\ead{lirh@ihep.ac.cn}

\credit{Writing, Simulation}

\address[1]{organization={Institute of High Energy Physics, Chinese Academy of Sciences},
    addressline={19B Yuquan Road, Shijingshan District}, 
    city={Beijing},
    postcode={100049},
    country={China}}

\author[1,2]{Yichen Li}[type=editor,
                        auid=000,bioid=1,
                        prefix=,
                        role=,
                        orcid=0000-0003-3042-0893]

\cormark[1]
\fnmark[1]

\ead{liyichen@ihep.ac.cn}


\credit{Writing, Simulation, Unfolding, Discussion}
                        
\author[1,2]{Zhimin Wang}[%
   role=,
   suffix=,
   orcid=0000-0002-8651-8999
   ]

\ead{wangzhm@ihep.ac.cn}

\credit{Coordination, Measurement}

\address[2]{organization={State Key Laboratory of Particle Detection and Electronics},
    addressline={19B Yuquan Road, Shijingshan District}, 
    city={Beijing},
    postcode={100049},
    country={China}}

\author[1,2,3]{Qiang Li}
\ead{qiangli@ihep.ac.cn}

\credit{Measurement}

\address[3]{organization={Spallation Neutron Source Science Center},
    addressline={},
    city={Dongguan},
    postcode={523803},
    country={China}}

\author[1,2]{Liang Zhan}[%
   role=,
   suffix=,
   ]

\ead{zhanl@ihep.ac.cn}

\credit{Coordination, Discussion}

\author[1,2]{Jun Cao}[%
   role=,
   suffix=,
   ]

\ead{caoj@ihep.ac.cn}

\credit{Coordination, Discussion}

\cortext[cor1]{Corresponding author}




\begin{abstract}
The Taishan Antineutrino Observatory (TAO) is a ton-level liquid scintillator detector to be placed at 30\,m from a core of the Taishan Nuclear Power Plant for precise reactor antineutrino spectrum measurements. One important background for TAO physics are the interactions of ambient neutrons that can penetrate its outer shieldings. The neutrons fluence and energy spectrum are measured with a Bonner sphere spectrometer. 
Data is unfolded with the iterative Maximum-Likelihood Expectation–Maximization (MLEM) method.
A simulation based on Geant4 is performed to provide the initial input spectrum to the unfolding and to understand the unfolded result.
The total neutron fluence rate is measured to be 36.1 $\pm$ 4.7 $Hz/m^2$, which is higher than the expectation.
For neutrons with kinetic energy lower than 20\,MeV, the measured fluence rate and energy spectrum can be well reproduced by simulation. While for the region greater than 20\,MeV, a significant discrepancy is observed and shall be understood with further studies.
\end{abstract}



\begin{highlights}
\item Fluence rate and energy spectrum of the ambient neutron at the Taishan Antineutrino Observatory (TAO) experiment site is measured and found to be in good agreement with Geant4 simulation for neutrons with kinetic energy less than 20\,MeV.
\end{highlights}

\maketitle
\flushbottom

\section{Introduction}
\label{sec:1:intro}

The Taishan Antineutrino Observatory (TAO\cite{TAO-CDR}) is a satellite experiment of the Jiangmen Underground Neutrino Observatory (JUNO\cite{JUNO-CDR,JUNO-yellow-book-2016,JUNO-detector}). It mainly consists of a ton-level liquid scintillator detector with sub-percent energy resolution (better than 2\% @1MeV) and will be placed in a basement in the vicinity of one core of the Taishan Nuclear Power Plant in Guangdong, China. The main purposes of the TAO experiment are 1) to provide a reference reactor antineutrino energy spectrum for JUNO; 2) to provide a benchmark measurement to test nuclear databases; 3) to search for light sterile neutrinos with a mass scale around 1\,eV; 4) to provide increased reliability and verification of the technology for reactor monitoring and safeguard. 

Located in a basement with a floor height of -9.6 meters, the effective overburden of TAO corresponds to roughly 4 meters of concrete. With such a shallow overburden, neutrons induced by cosmic ray muons interacting with the surroundings, such as the concrete floors in the building and the steel cable bridge in the basement, are the dominant component of the ambient neutron, which can form a potential background to the TAO experiment especially when the parent muon cannot be tagged by TAO's veto system. 
Therefore, ambient neutron production has to be measured to guide the neutron shielding design.
Other sources of neutrons include cosmic ray neutrons penetrating the overburden and radiogenic neutrons from ($\alpha$,n) processes, both of which are negligible compared to the cosmic ray muon-induced neutrons. 

Underground ambient neutron were measured for various depths and profiles of overburden in several short-baseline reactor anti-neutrino experiments, such as 
DANSS\cite{DANSS-ALEKSEEV201856}, Solid\cite{solid-Abreu_2017}, MINER\cite{MINER-AGNOLET201753}, NEUTRINO-4\cite{neutrino-4-PhysRevD.104.032003}, PROSPECT\cite{PROSPECT-ASHENFELTER2019287}, NEOS\cite{NEOS-PhysRevLett.118.121802}, and in several other underground laboratories \cite{neutron-bkg-HASHEMINEZHAD1995524,CJPL-neutron-ZENG2015108,n-modan-CHAZAL1998163,Manukovsky2016631,Schmidt201328,Boliev2017,Persiani2013235,Blyth2016,Aharmim2019,Trzaska2019,Niese2007173}. 
Models for the yield and spectrum of the cosmic ray muon-induced neutrons were developed \cite{Agafonova2013607,Malgin2017,Malgin2017728} and dedicated experiments exist \cite{Abt20171,Kneibetal201987}
Overall, good agreements between data and model has been achieved. But not all measurements were well documented and are compatible with each other. Not all were supplemented with simulation and can be well matched with simulation or calculation \cite{LINDOTE2009366}. Few results can be found at the shallow overburden of $\sim$10 m.w.e. similar to TAO. As commented in Reference\,\cite{PhysRevD.101.123027}, the current situation of the underground ambient neutron study is not very satisfactory. 

In this paper, dedicated measurements of ambient neutrons interactions in the basement where TAO will be positioned are reported. The ambient neutron fluence and energy spectrum are measured using a Bonner sphere spectrometer\,\cite{bonner-review-THOMAS200212}. A thorough Geant4\,\cite{geant4-AGOSTINELLI2003250} simulation has been done to understand the result as well as to provide an initial spectrum to the unfolding of the energy spectrum. The aims of this paper include 1) to facilitate neutron shielding design of the current and possible future experiments (such as Liquid Argon and Gas TPC) to be performed in the same basement; 2) to validate the Geant4 simulation which is the basis of the TAO neutron shielding design; 3) to provide a new reference that deals with shallow overburden underground ambient neutrons, the literature of which is limited (not to mention those supplemented with detailed simulation to understand the results); 4) to provide new reference illustrating the details of the Monte Carlo simulation of underground ambient neutrons and discussing the impact of tuning the Monte Carlo on the resulting neutron fluence and energy spectrum, which is not always done in other literature.

The structure of the paper is as follows. Section\,\ref{sec:1:lab} introduces the TAO laboratory with more details. The measurement with Bonner sphere spectrometer is described in Section\,\ref{sec:1:meas}, with the unfolding of the collected Bonner sphere counts explained in Section\,\ref{sec:1:unfold}. The simulation of the neutron fluence and energy spectrum and their comparison with the measured results are presented in Section\,\ref{sec:1:simulation}. Section\,\ref{sec:1:sum} gives a summary.

\section{Taishan Antineutrino Observatory}
\label{sec:1:lab}

\begin{figure}[!htb]
    \centering
	\includegraphics[width=0.75\linewidth]{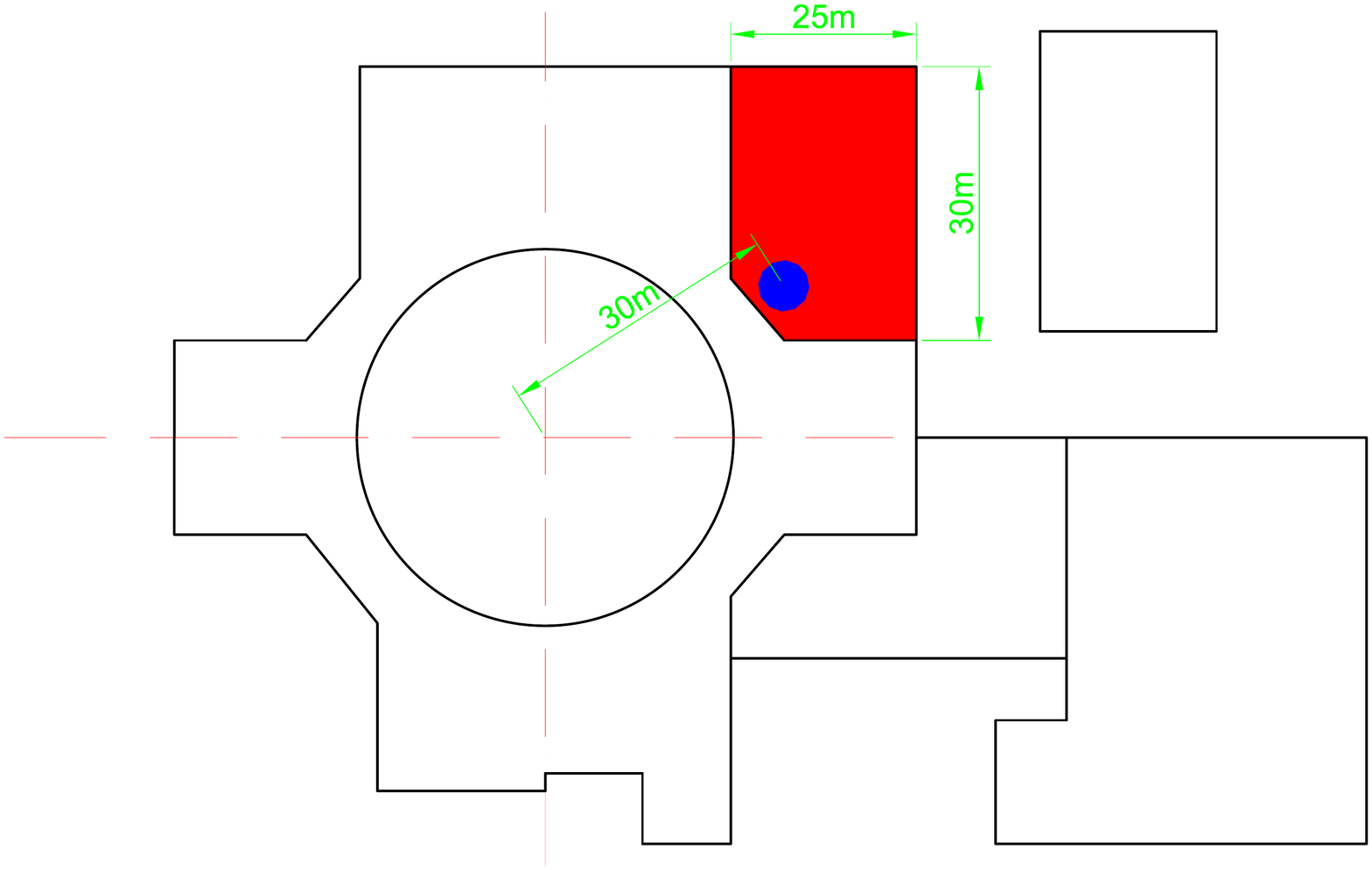}
    \caption{Technical drawing of the Taishan Nuclear Power Plant core and TAO detector. The circle represents the core containment. The basement where TAO will be located is shown in red. The liquid scintillator detector will be placed close to the cutting corner of the basement.}
    \label{fig:tao-lab}
\end{figure}

The TAO experiment will be located at the Taishan Nuclear Power Plant, in a basement at -9.6\,m underground,  outside of the concrete containment shell, highlighted in red in Figure\,\ref{fig:tao-lab}. The dimension of the basement is approximately 30(L)x25(W)x5(H) $m^3$ (ignoring the cutting corner). The TAO liquid scintillator detector will be placed close to the wall of the cutting corner of the basement, which makes it as close to the reactor core as possible. The horizontal distance from the center of the detector to that of the reactor core is around 30 meters.

The overburden of the basement is mainly composed of the concrete floors above the basement and the roof of the building. From the drawings provided by the nuclear power plant, the overburden is not uniform. In most cases, the total thickness of the concrete exceeds 3 meters, while in some regions, it is only 2.5 meters.
On the other hand, the effective overburden is estimated to be around 4 meters by a muon rate ratio measurement between the ground and underground. 
Although there is uncertainty in the exact thickness of the overburden, it will be shown by simulation in Section \ref{sec:1:simulation} that the ambient neutron is barely affected.

\begin{figure}[!htb]
    \centering
    \begin{subfigure}{a}
    \includegraphics[width=0.30\linewidth]{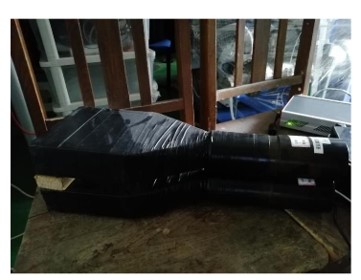}    
    \end{subfigure}
	\begin{subfigure}{b}
	\includegraphics[width=0.5\linewidth]{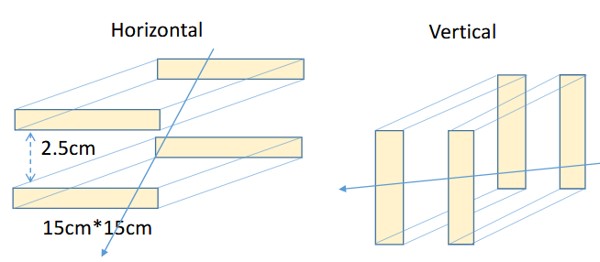}    
	\end{subfigure}
    \caption{The photo (a) and layout (b) of the plastic scintillator modules used to perform the muon rate measurement. The material between the two plastic scintillator tiles is a wooden spacer with 2cm thickness.}
    \label{fig:muon-measurements}
\end{figure}

In the muon rate ratio measurement, the muons are tagged by the coincidence of two plastic scintillator modules in parallel as shown in Figure\,\ref{fig:muon-measurements} and the results are summarized in Table\,\ref{tab:muon-meas}. The plastic scintillator modules are horizontally and vertically placed respectively to measure muon rates in different directions. A Geant4 simulation is performed to simulate the muons that pass through the overburden. The thickness of the concrete in the simulation is tuned to 4 meters so that the simulated muon rate ratios between the ground and underground are in good agreement with the measured ones. There is further room for tuning to make the agreement better, but it is not necessary since the resulting change is much smaller than the uncertainty of the measurement in this work.

\begin{table}[!htb]
    \centering
    \caption{The measured rates of muons passing through both the plastic scintillator modules on the ground and underground in the horizontal and vertical arrangements of the modules. The numbers are given without area normalization. The rates corresponds The Geant4 simulated rates correspond to 4 meters of overburden. The statistical error is shown for data, while it is negligible for simulation.}
    \label{tab:muon-meas}
    \begin{tabular}{l|c|c|c|c|c|c}
    \hline
    \hline
     & \multicolumn{3}{c|}{Measurement} & \multicolumn{3}{c}{Simulation}  \\
    \hline
    Rate (Hz) & Above ground & Underground & Ratio & Above ground & Underground & Ratio   \\
    \hline
    Horizontal & 2.32$\pm$0.09 & 0.69$\pm$0.04 & 0.30$\pm$0.02 & 2.30 & 0.79 & 0.34   \\
    \hline
    Vertical & 0.79$\pm$0.05 & 0.20$\pm$0.02 & 0.25$\pm$0.03 & 0.76 & 0.19 & 0.30 \\
    \hline
    \hline
    \end{tabular}
\end{table}

\begin{figure}[!htb]
    \centering
	\includegraphics[width=0.6\linewidth]{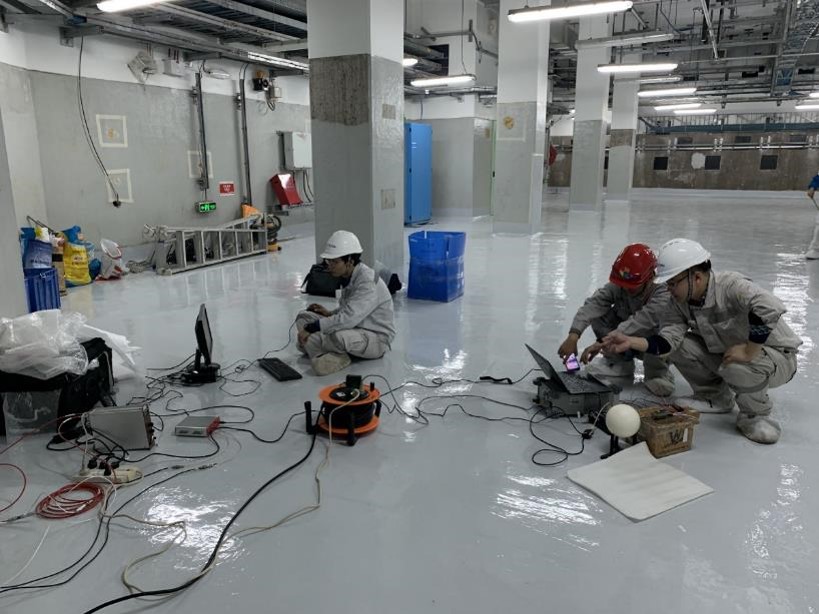}
    \caption{The interior of the basement where TAO will be located. Cable trays along the wall and gas and water pipes on the ceiling can be seen.}
    \label{fig:lab-view}
\end{figure}

Figure\,\ref{fig:lab-view} shows the TAO experimental area. This photograph was taken at the point where the TAO detector will be located. There are long and multi-layer cable trays along the wall, and beam and water pipes on the ceiling. People in the photo were performing the muon rate measurement with the plastic scintillator modules and the ambient neutron measurement with a Bonner sphere spectrometer. 

\section{Bonner sphere spectrometer measurements}
\label{sec:1:meas}

The Bonner sphere spectrometer (BSS) is widely used for ambient neutron measurement for its wide energy coverage and isotropic response. BSS is usually composed of 8-12 polyethylene spheres with a neutron-sensitive detector in the center. They are designed in such a way that the response functions over neutron energy are quite different so that their count rates under a specific neutron environment are different. Usually, the sphere diameters are 3 to 12 inches, and the neutron response function ranges from thermal neutron energy to about 20 MeV. To expand to higher energy regions, extended Bonner spheres with lead shell sandwiched between the polyethylene shells are often used. The lead-containing spheres can extend the neutron sensitivity above 1 GeV. 

\begin{figure}[!htb]
    \centering	
    \begin{subfigure}{a}
    \includegraphics[width=0.36\textwidth]{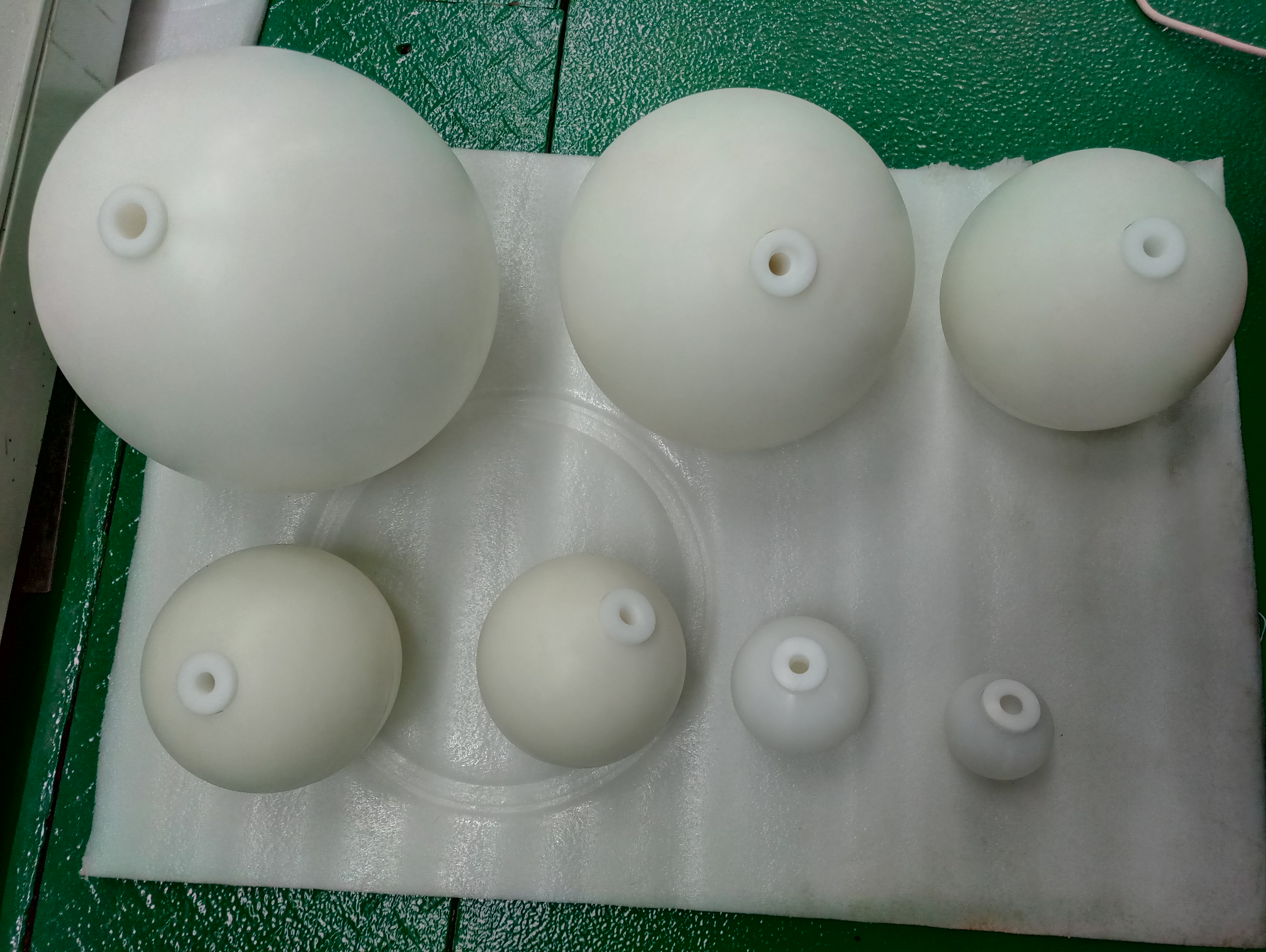}
    \end{subfigure}
	\begin{subfigure}{b}
	\includegraphics[width=0.36\linewidth]{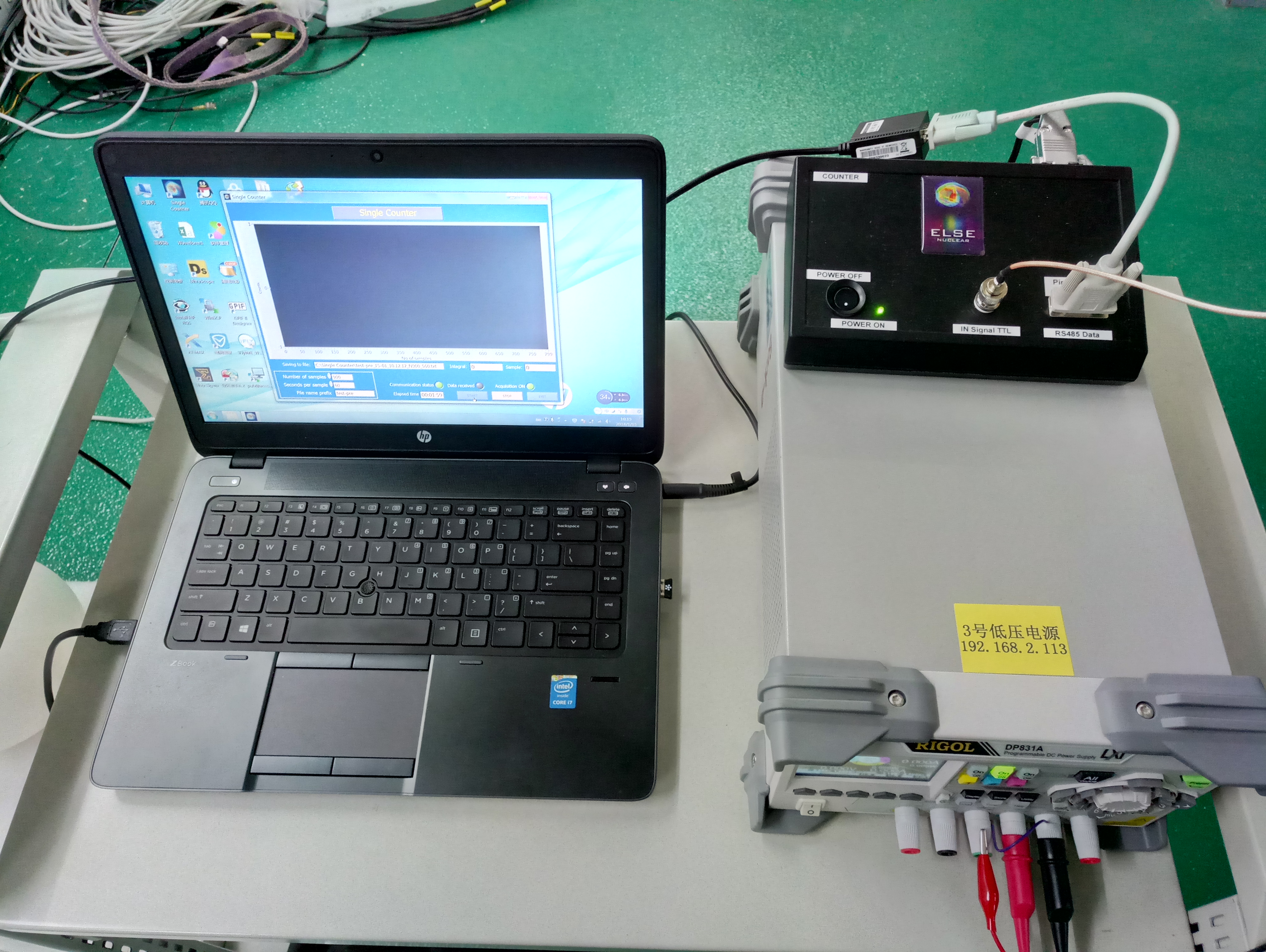}
    \end{subfigure}
    \caption{The Bonner sphere spectrometer used in the measurement (a) and the associated electronics and power modules (b).}
    \label{fig:BSSphoto}
\end{figure}

\begin{figure}[!htb]
    \centering
	\includegraphics[width=0.95\linewidth]{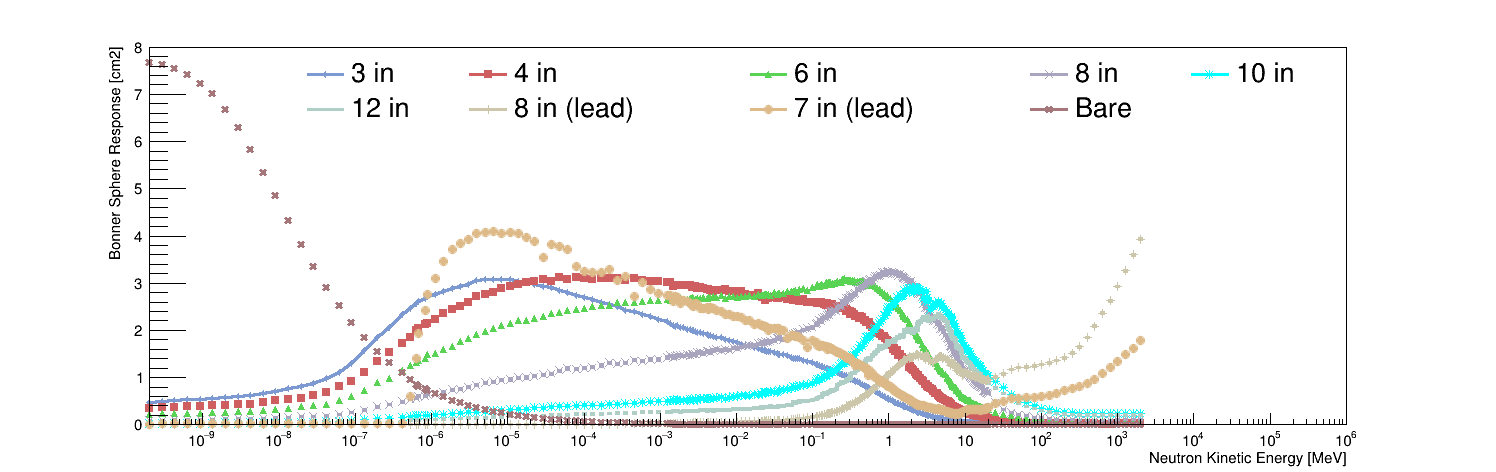}
    \caption{The response functions of the Bonner sphere spectrometer.}
    \label{fig:BS-response}
\end{figure}

The BSS used in this measurement (Figure\,\ref{fig:BSSphoto}) was purchased from the Else Nuclear Ltd., Italy, and is composed of six polyethylene spheres with diameters of 3, 4, 6, 8, 10, 12 inches and two lead-containing modified multi-shell spheres with diameters of 7 and 8 inches. The central neutron detector is a spherical ${He}^3$ proportional counter SP9 purchased from the Centronic Ltd., UK, with a pressure of 1.2 atm, which works individually with the eight spheres. The neutron response functions of the Bonner spheres, defined as the reading per unit neutron fluence (in unit of $cm^2$) in the energy range from 10$^{-3}$\,eV to 1\,GeV, are provided by the vendor using simulation, as shown in Figure\,\ref{fig:BS-response}. The BSS response functions were validated with 144 keV and 1.2 MeV neutron sources at the China Institute of Atomic Energy, and the deviation obtained is less than 5\%. An integrated electronics module is used to supply 1200 V high voltage to the ${He}^3$ detector and to collect count rates. Before performing the measurement at Taishan, the voltage and gain of the electronics were appropriately adjusted, resulting in an overall scaling factor of 1/2.25 on the response functions (not included in Figure\,\ref{fig:BS-response}), with a 10\% uncertainty. The BSS has been used to measure the natural environment on the ground in Dongguan City, China, and the neutron background in experimental halls of Back-n at CSNS, and the results are consistent with simulations and other literature, verifying the good performance of the spectrometer \cite{LI2020164506}.

The measurements were performed at the location shown in Figure\,\ref{fig:lab-view} before the reactor was turned on. Due to the limited experimental time in the sensitive area of the nuclear power plant, each sphere was placed there for an average duration of 3 hours to accumulate neutron counts (durations of individual spheres vary due to the actual conditions). Besides, an additional measurement with the bare $He^3$ detector was performed to gain sensitivity to thermal neutrons. The measured results are summarized in Table\,\ref{tab:BS-meas}. Due to the very low ambient neutron rate in the experimental hall, the cumulative counts for different spheres are only a few to dozens. Such low statistics will bring significant statistical uncertainty to the unfolded neutron spectrum to be shown in Section\,\ref{sec:1:unfold}. It is planned to redo the measurement in the future with a longer duration to gain enough statistics for a precise spectrum measurement.

\begin{table}[!htb]
    \centering
    \caption{The raw count, duration, and count rate of each Bonner sphere.}
    \label{tab:BS-meas}
    \begin{tabular}{l|c|c|c|c|c|c|c|c|c}
    \hline
    \hline
    Bonner Sphere & Bare & 3 in & 4 in & 6 in & 8 in & 10 in & 12 in & 8 in (lead) & 7 in (lead) \\
    \hline
    Raw Count & 5 & 12 & 17 & 25 & 13 & 9 & 27 & 12 & 22 \\
    \hline
    Duration (h) & 1 & \multicolumn{5}{c|}{3} & 10 & \multicolumn{2}{c}{3} \\
    \hline
    Rate (cph) & 5 & 4 & 5.7 & 8.3 & 4.3 & 3 & 2.7 & 4 & 7.3 \\
    \hline
    \hline
    \end{tabular}
\end{table}

\section{Unfolding}
\label{sec:1:unfold}

The unfolding of the neutron differential fluence rate in kinetic energy, i.e. the energy spectrum, consists in solving a group of equations in which the convolution of the Bonner sphere response functions with the energy spectrum is equal to the measured count rate for every sphere. It is challenging and most of the time suffers from ambiguity. This is because the number of bins in the neutron energy spectrum is usually much more than the number of measurements, making the equation group underdetermined and having multiple outcomes.
A recent international comparison exercise on neutron spectra unfolding in Bonner spheres spectrometry can be found in Reference\,\cite{GOMEZROS2022106755}, where a variety of unfolding methods are described.
An iterative method is usually applied and a prior spectrum (initial guess spectrum) as close as possible to the true spectrum is required. 
In choosing the initial spectrum and the number of iterations, empirical knowledge is often needed.
In another method, the spectrum to be unfolded is parametrized into a combination of 3 or 4 analytical functions with just a few parameters (equal to or less than the number of Bonner spheres). Then a unique spectrum can be fitted, but the price is a potential bias due to the modeling of the true spectrum by the chosen functions.

In this paper, an iterative Maximum-Likelihood Expectation–Maximization (MLEM) method is applied, the code of which is shared by McGill University on GitHub \cite{MONTGOMERY2020163400} and adapted for this study. The MLEM method is shortly illustrated here. Having defined the measured count rate of a sphere as $C_i$, then it can be expressed as the convolution of the neutron energy spectrum $\Phi_j$ and the Bonner sphere response functions $R_{ij}$:
\begin{equation}
    C_i = \sum_{j} R_{ij} \times \Phi_j
\label{eqn:conv}
\end{equation}
where $i$ is the sphere index and $j$ the energy bin index. 
At iteration $k$, the energy spectrum in the previous iteration (or the initial spectrum if it is the first iteration) is updated by scaling each bin of the spectrum by the normalized ratio of the raw count rates to the reconstructed count rates defined by Equation\,\ref{eqn:ratio}:
\begin{equation}
    \Phi_j^{k+1} = \Phi_j^k \times \frac{\sum_{i} R_{ij} \frac{C_i}{\sum_{b} R_{ib} \Phi_b^k}}{\sum_i R_{ij}}
\label{eqn:ratio}
\end{equation}
where $b$ is also the energy bin index.

\begin{figure}[!htb]
    \centering
	\includegraphics[width=0.6\linewidth]{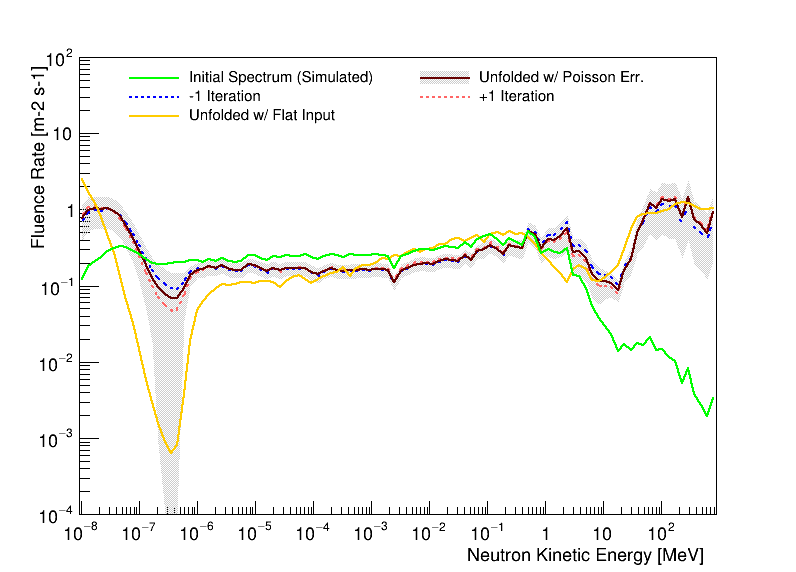}
    \caption{The unfolded neutron energy spectrum compared with the Geant4 simulated initial spectrum. The shady error band represents the Poisson statistical uncertainty. The number of iterations is increased or decreased by 1 to check the corresponding systematic effect. The unfolded spectrum starting from a flat initial spectrum is shown.}
    \label{fig:spectrum}
\end{figure}

The initial spectrum for the unfolding is important. If it deviates from the underlying true spectrum too much, there is a large chance to get a wrong unfolded spectrum. In this study, the initial spectrum is taken from the Geant4 simulation which will be described in Section\,\ref{sec:1:simulation} and shown in lethargy representation as the green histogram in Figure\,\ref{fig:spectrum}. The binning of the unfolded spectrum is determined by the input initial spectrum: starting from 10$^{-8}$\,MeV to around 800\,MeV, and divided into 100 bins with a constant bin width on a log scale. A flat initial spectrum is also tried to study the systematic uncertainty of the choice of the input spectrum.

\begin{figure}[!htb]
    \centering
	\includegraphics[width=0.6\linewidth]{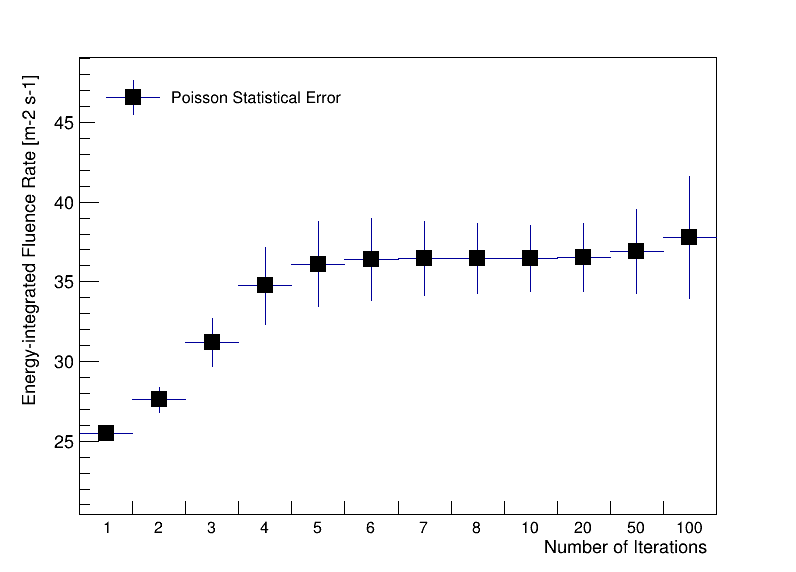}
    \caption{The evolution of the energy-integrated neutron fluence rate as a function of the number of iterations of the MLEM method.}
    \label{fig:fluxes}
\end{figure}

The MLEM method has to be stopped at certain iterations, otherwise, once it passes optimal point, it will just acquire more and more noise as the number of iterations further increases. Here the algorithm is stopped when the energy-integrated fluence rate converges, as shown in Figure\,\ref{fig:fluxes}, where 5 iterations is found to be a proper choice. The number of iterations is varied up and down by 1 to study the corresponding systematic uncertainty.

The statistical uncertainty due to the fluctuation of the recorded Bonner sphere counts is studied with a simple method. A data set is generated by sampling raw counts from the Poisson distributions with mean values set to the observed counts of each sphere respectively. By calculating the covariance matrix of the unfolded toy spectra, the statistical uncertainty can be evaluated.
Another systematic effect considered is the uncertainty of the universal scaling factor applied to the Bonner sphere response functions mentioned in Section\,\ref{sec:1:meas}, which is taken to be 10\%.
The Monte Carlo statistical uncertainty of the simulated Bonner sphere response functions is studied and found to be negligible.

\input{tabs/fluxes.tex}

After 5 iterations, the energy-integrated fluence rate in the full range is 36.1 $\pm$ 2.8 (stat.) $\pm$ 3.9 (sys.) $Hz/m^2$ as shown in Table\,\ref{tab:fluxes}. The systematic uncertainty is dominated by the universal Bonner sphere scaling factor uncertainty. The fluence rate shows no dependency on the choice of the initial input spectrum. The simulated fluence rate is 21.6 $Hz/m^2$, which is 3 sigmas smaller than the unfolded rate.
To further clarify this deviation, the fluence rates integrated in 4 distinct energy regions, i.e. the thermal peak, the epithermal, the fast peak, and the high energy peak regions, are shown in Table\,\ref{tab:fluxes}. The most significant deviation comes from the high-energy neutron region ($E > 20$ MeV). Having excluded this region, the overall fluence rates are compatible between the measurement and the simulation. Another local deviation is observed in the relative strength between the thermal peak and epithermal regions: the data shows a much stronger thermal peak than the simulated. Both the deviations will be discussed in Section\,\ref{sec:1:simulation}. 

\begin{figure}[!htb]
    \centering
	\includegraphics[width=0.6\linewidth]{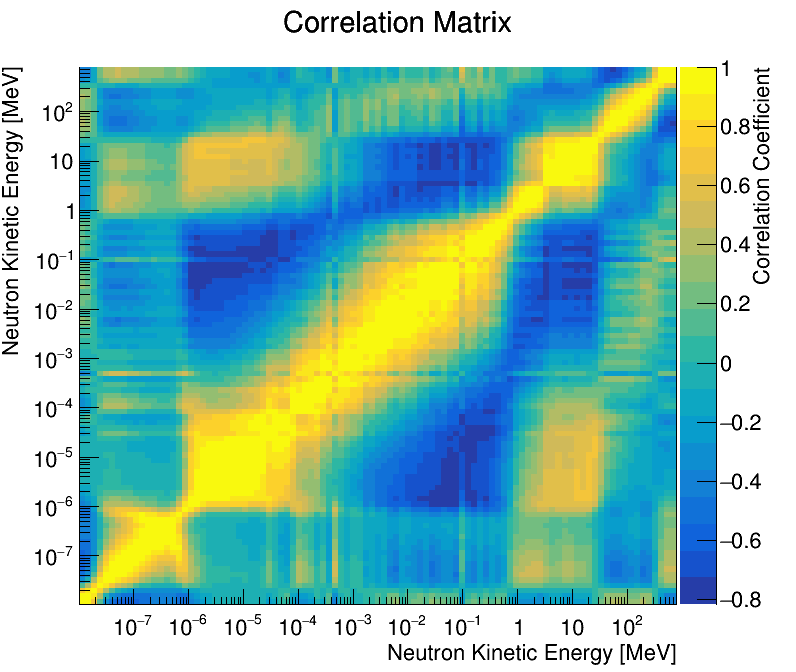}
    \caption{The statistical correlation matrix of each bin of the unfolded neutron energy spectrum.}
    \label{fig:corr}
\end{figure}

The unfolded energy spectrum is shown in Figure\,\ref{fig:spectrum}. The four regions mentioned above can be distinguished and the two deviations between the unfolded and simulated spectra are also identified. The tiny structures in the spectrum are inherited from the simulation, while the overall magnitudes of each region are determined by data. By comparing the unfolded and the initial spectra, it can be seen how the data bend the initial spectrum in blocks. The intersections between the thermal and epithermal regions and the magnitude of the high energy peak region are quite sensitive to Poisson fluctuation of the Bonner sphere counts. The unfolded spectrum is more or less stable against the variation of the number of iterations, which means a reasonable shape convergence. The unfolded spectrum starting from a flat initial spectrum is also shown in Figure\,\ref{fig:spectrum}. It takes more iterations to become stable, which is reasonable since the emergence of the four regions would cost some iterations while in the case of using the Geant4 spectrum as input, these regions are predefined. The difference between the two unfolded spectra is significant and can be taken as a conservative estimate of the spectrum uncertainty due to the choice of the initial spectrum.
The statistical correlation matrix of each bin is shown in Figure\,\ref{fig:corr}. The positive and negative correlations gather in blocks as expected, and the boundaries between the four energy regions can be identified.

\begin{figure}[!htb]
    \centering
	\includegraphics[width=0.6\linewidth]{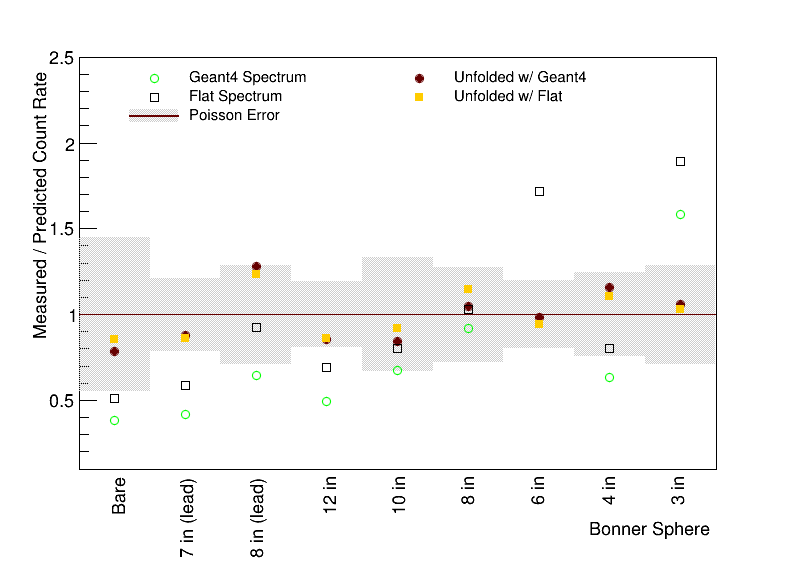}
    \caption{The count rates derived from the convolution of the Bonner sphere response functions with the Geant4 spectrum, the flat spectrum and their respective unfolded spectra, compared with the measured count rates. Before the convolution, the Geant4 and flat spectra are normalized to their corresponding unfolded spectra.}
    \label{fig:foldback}
\end{figure}

As a final check, the Geant4 simulated spectrum and flat spectrum as well as their respective unfolded spectra are convoluted with the Bonner sphere response functions to derive a set of estimated Bonner sphere count rates and compared with the measured rates, as shown in Figure\,\ref{fig:foldback}. To focus on shape comparison, the Geant4 and flat initial spectra are normalized to their corresponding unfolded spectra. As can be seen, the initial spectra predict a set of count rates way off the measured rates, while after unfolding, the predicted rates become compatible with the measured rates within their Poisson statistical uncertainty. 

\section{Geant4 Simulation}
\label{sec:1:simulation}

For further understanding and comparison with the measurements as well as providing an initial spectrum to the unfolding, a standalone project based on Geant4 (version 10.4.2) has been developed to simulate the muon-induced ambient neutron in the basement. It is also an attempt to give as detailed as possible description of the Monte Carlo simulation of the underground ambient neutron in shallow overburden.

\begin{figure}[!htb]
    \centering
	\begin{subfigure}{a}
    \includegraphics[width=0.342\textwidth]{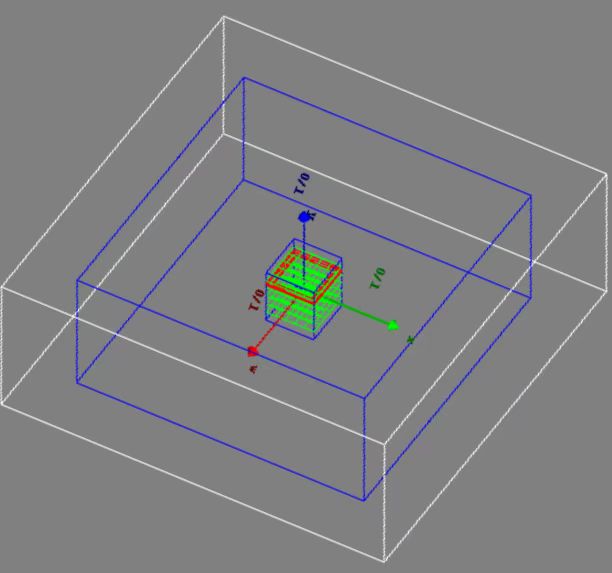}
    \end{subfigure}
	\begin{subfigure}{b}
    \includegraphics[width=0.36\textwidth]{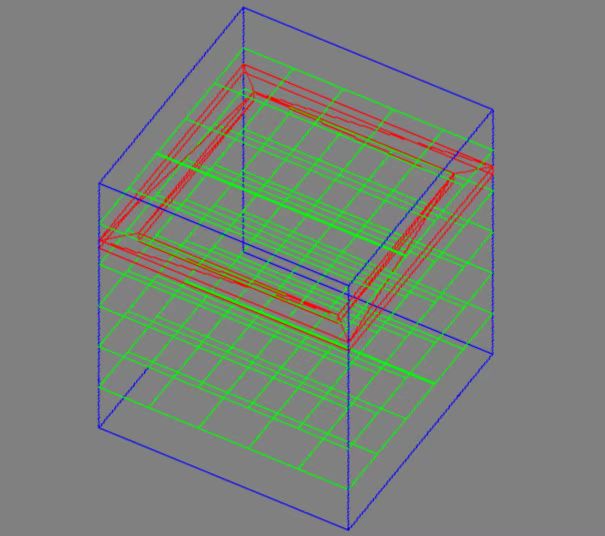}
    \end{subfigure}
    \caption{The Geant4 simulation geometry (a), including the world (white box, 40(L)x40(W)x(15) $m^3$), the concrete (big blue box, 30(L)x30(W)x13(H) $m^3$), the basement (small blue box, 5(L)x5(W)x6(H) $m^3$), the steel cable tray on the wall of the basement (red line), and the 5x5x5 matrix of virtual plane detectors to count neutron fluence (green). The zoomed basement (b) is shown.}
    \label{fig:geo}
\end{figure}

In the Geant4 simulation, the geometry of the basement is simplified as shown in Figure\,\ref{fig:geo}. A small blue box of dimension 5(L)x5(W)x6(H) $m^3$, representing the basement, is located in the center of a chunk of concrete, represented by a big blue box of dimension 30(L)x30(W)x13(H) $m^3$. The concrete is configured to uniformly contain water with a mass fraction of 0.4\%, according to Reference\,\cite{WaterContent}, and is located in a world of dimension 40(L)x40(W)x(15) $m^3$. On the wall of the basement, there is a belt of steel cable tray, the shape of which is represented by red lines. In the space of the basement, a 5x5x5 matrix of virtual plane detectors (in green) is deployed horizontally, each having a sensitive area of 1 $m^2$. Whenever a neutron passes through any one of the detectors from any direction, it will record the angle of incidence $\theta$ on the plane it enters and the ID of the detector through which it penetrated. Each passage of a neutron adds to the counting of fluence by 1/$\cos{\theta}$. A right-hand Cartesian coordinate system is illustrated in the figure, with the $z$-axis upward.

\begin{figure}[!htb]
    \centering
    \begin{subfigure}{a}
    \includegraphics[width=0.45\textwidth]{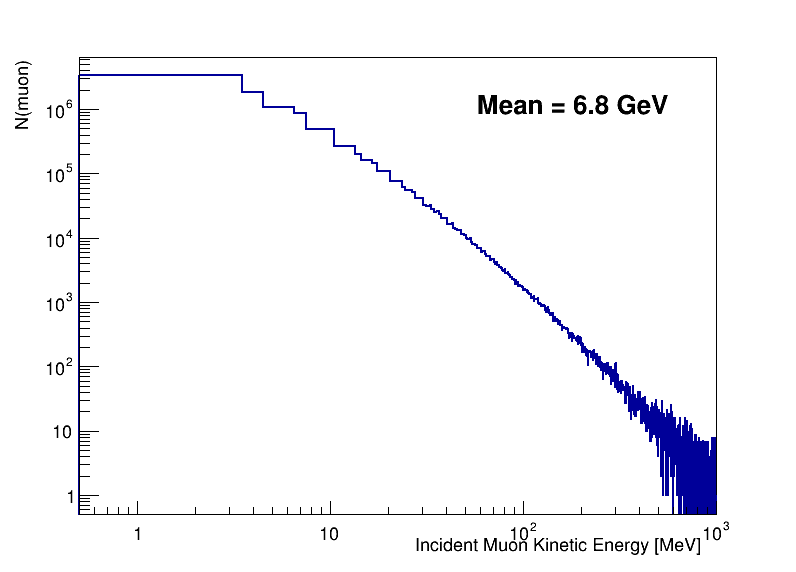}
    \end{subfigure}
	\begin{subfigure}{b}
    \includegraphics[width=0.45\textwidth]{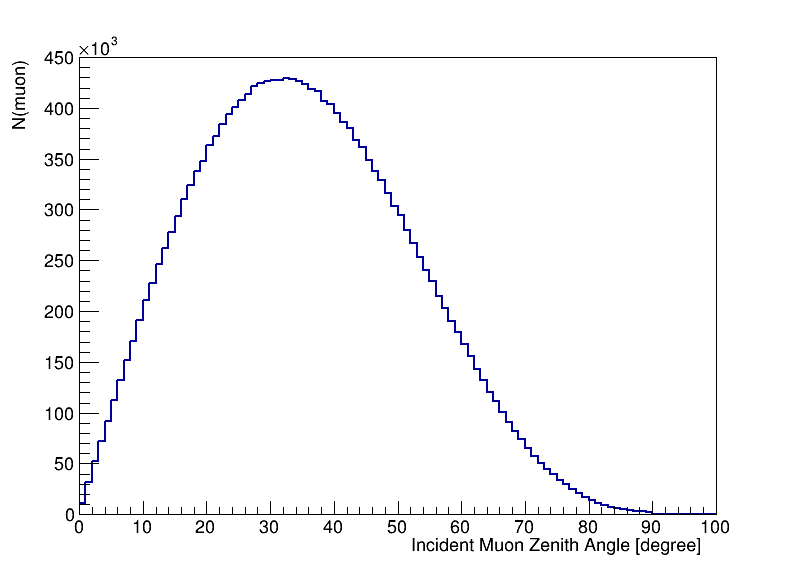}
    \end{subfigure}
    \caption{The distributions of the incident muon kinetic energy (a) and zenith angle (b) in the simulation.}
    \label{fig:initmuon}
\end{figure}

A sea level cosmic ray muon sample (including both muon and antimuon) of $2 \cdot 10^7$ events is generated uniformly on the surface of the concrete box over an area of 30x30 $m^2$, with a rate of 167 $Hz/m^2$ \cite{muon-rate}. The energy and incident angle distributions of the sample are derived from \cite{muon-energy}, which are shown in Figure\,\ref{fig:initmuon}. 
When a muon passes through the concrete, processes such as muon capture, muon nuclear interaction, photo nuclear interaction, and electromagnetic and hadronic cascades can happen, through which neutrons are produced with different probabilities and kinetic distributions.
About one-third of the muons are captured by the elements in the concrete, resulting in muon-capture induced neutrons, which are the dominating contribution to the ambient neutron in the basement. Another one-third of the muons (which are actually antimuon) decay at rest after they dissipated their kinetic energy in the concrete and the last one-third passes through the geometry. The latter two types of the muons generate neutrons mainly via muon and photo nuclear processes, yielding less neutrons than that of the muon capture.

\begin{figure}[!htb]
    \centering
    \begin{subfigure}{a}
    \includegraphics[width=0.45\textwidth]{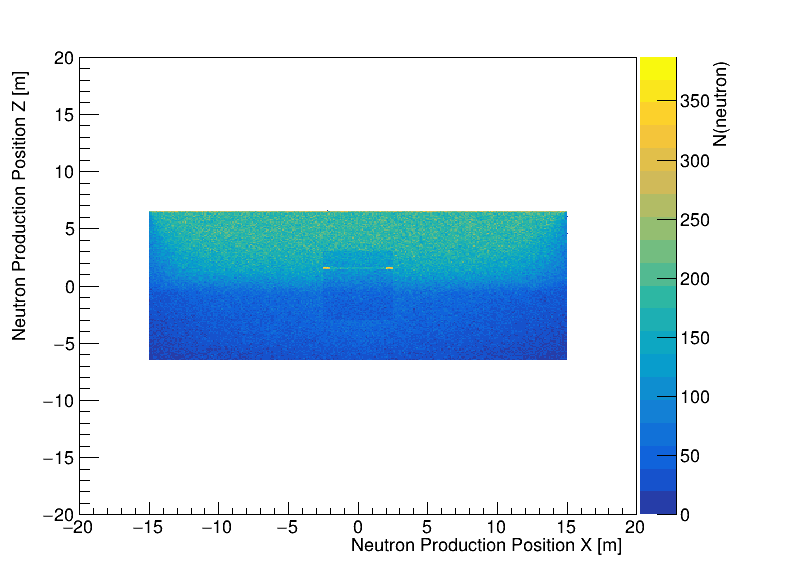}
    \end{subfigure}
	\begin{subfigure}{b}
    \includegraphics[width=0.45\textwidth]{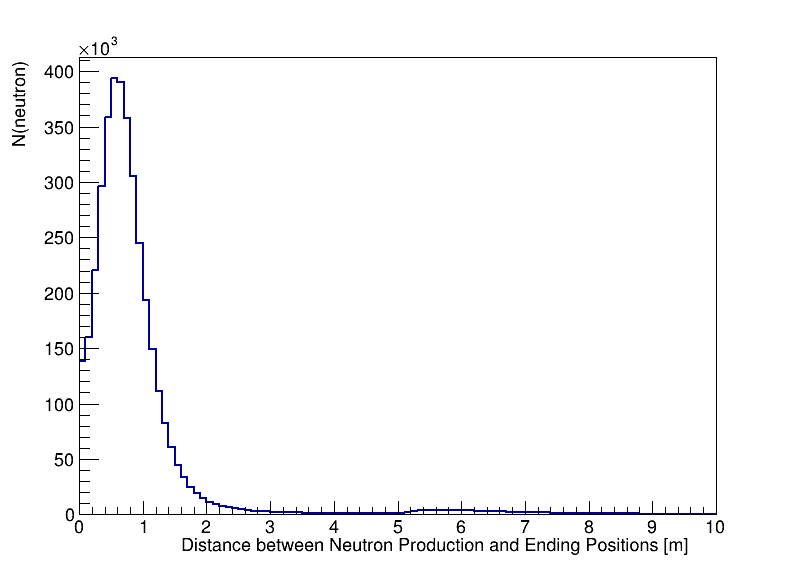}
    \end{subfigure}
    \caption{The distributions of neutron production position in the (X, Z) plane (a) and the distance between the neutron production and ending positions (b).}
    \label{fig:neutron1}
\end{figure}

In the geometry considered, most of the neutrons are generated from the muon capture process and end up being captured by the elements in the concrete or exiting the geometry. The distribution of the neutron production position in the (X, Z) plane is shown in Figure\,\ref{fig:neutron1} a. As it goes deeper into the concrete downward, less neutrons are generated. This is the result of the falling kinetic energy distribution of the incident muons. In Figure\,\ref{fig:neutron1} b, the distribution of the distance between the production and ending positions of the neutrons is shown. Most of the neutrons got captured within 2 meters from their production vertices. The long tail of the distribution is not fully due to the high penetrating power of the high-energy neutrons. Geometric effect also contributes, such as the hollow basement and the neutrons flying out of the geometry without being captured, both making the neutron ending point further away from the corresponding production vertex. 

\begin{figure}[!htb]
    \centering
    \begin{subfigure}{a}
    \includegraphics[width=0.45\textwidth]{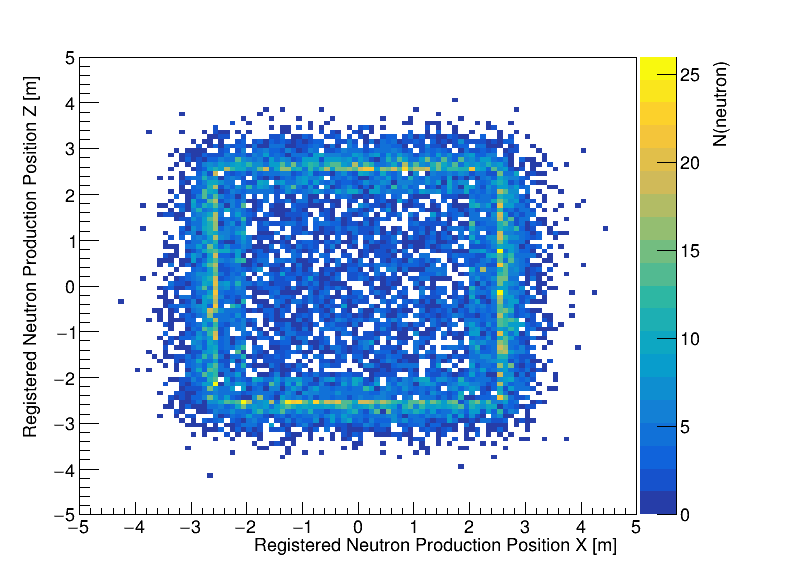}
    \end{subfigure}
	\begin{subfigure}{b}
    \includegraphics[width=0.45\textwidth]{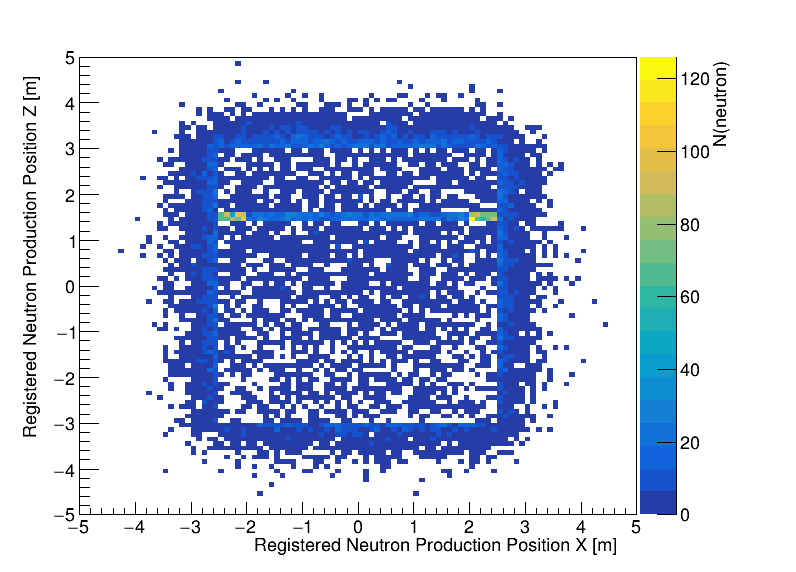}
    \end{subfigure}
    \caption{The distributions of production vertices in the (X,Y) (a) and (X,Z) (b) planes for neutrons that are registered by the virtual detector array in the basement.}
    \label{fig:neutron2}
\end{figure}

Not all neutrons produced in the geometry can reach the interior of the basement. In Figure\,\ref{fig:neutron2}, the production vertices of the neutrons registered by the 5x5x5 matrix of virtual plane detectors placed in the basement are shown in the (X, Y) and (X, Z) planes. As can be seen, only the neutrons generated within roughly 1 meter off the wall of the basement can reach the detector array, thus contributing to the counting of neutron fluence. Besides, the shapes of the wall and the steel cable tray on the wall can be seen as hot regions. After averaging over the 125 detectors, the simulated mean total fluence rate is 21.6 $Hz/m^2$, which is way off the measured total fluence rate presented in Section\,\ref{sec:1:meas}. A possible explanation of this discrepancy will be discussed below. The fluence rate distribution across the 125 detectors is rather uniform, with a slightly higher rate in the detectors close to the ceiling of the basement. This is expected, since more muons are captured in the shallow region of the concrete.

\begin{figure}[!htb]
    \centering
    \begin{subfigure}{a}
    \includegraphics[width=0.45\textwidth]{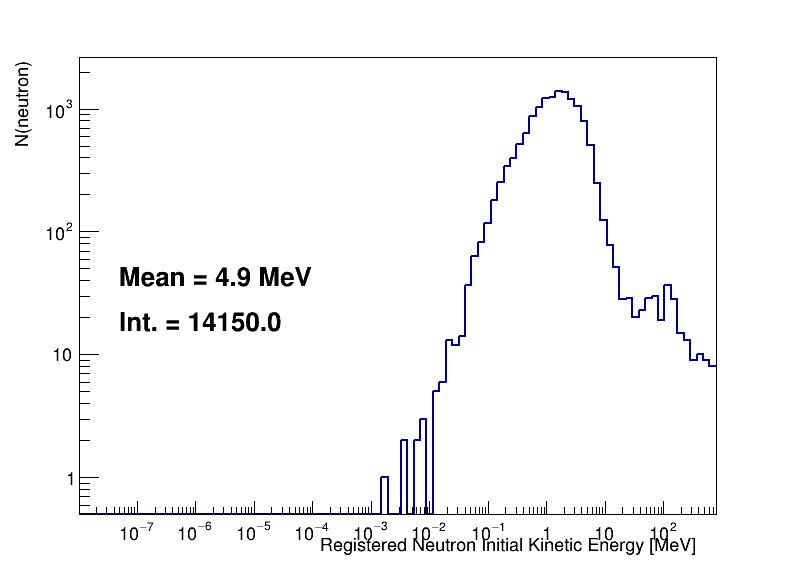}
    \end{subfigure}
	\begin{subfigure}{b}
    \includegraphics[width=0.45\textwidth]{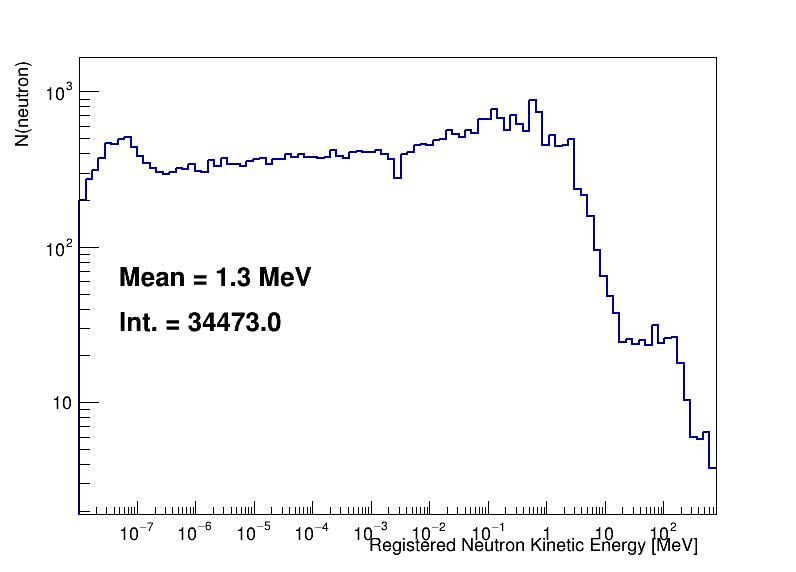}
    \end{subfigure}
    \caption{The distributions of the initial kinetic energy (a) and the transient kinetic energy at the moment of neutron registration (b) for the neutrons registered by the virtual detector array.}
    \label{fig:kinmodu}
\end{figure}

The concrete modulation effect on the kinetic energy distribution of the neutrons registered by the detector array can be seen in Figure\,\ref{fig:kinmodu}, where the distributions of the initial kinetic energy and the transient kinetic energy at the moment of the neutron registration are shown. After passing through the concrete wall and getting reflected back and forth between the walls, the initial energy spectrum of the neutrons is flattened down to the low energy region, and the thermal peak is formed. The modulated spectrum has statistics more than twice the initial spectrum, which is the result of multiple registrations of the same neutron. It will be shown later that this is not an issue and will not bias the calculation of neutron fluence rate.

After normalizing to the simulated mean total neutron fluence rate, the mean kinetic energy spectrum of the 125 detectors is used as initial input for the unfolding introduced in Section\,\ref{sec:1:unfold} and compared with the unfolded spectrum, as shown in Figure\,\ref{fig:spectrum}.
There are mainly two discrepancies between the simulated and unfolded spectra on the strength of the thermal neutron and high-energy neutron peaks respectively. 

The relatively lower thermal peak in the simulation than in data can be easily understood by the lack of accurate water content information in the concrete. In the simulation, a mass fraction of 0.4\% of water is added uniformly to the concrete. In reality, not only the mass fraction of water but also its distribution across different layers of the concrete can vary, which will change the magnitude of the thermal peak significantly.

The excessive high-energy neutron peak in data cannot be understood by the current simulation and is also not seen in relevant literature. If excluding the high energy peak, the simulated and measured total fluence rates can be well matched, as can be seen in Table\,\ref{tab:fluxes}. The spectrum can also be matched if the water content in the concrete is adjusted to a reasonable amount. Thus it is speculated that the main source of the observed high-energy neutron peak is not cosmic ray muon-induced neutron but something else. Although it is not clear what this source could be, simulation has been done to rule out the following possibilities: the surviving cosmic ray neutrons passing through the overburden; the secondary neutrons generated by cosmic ray proton interacting with the overburden. It is also wondered if the counts of the two lead-containing Bonner spheres are reliable since if they are excluded from the unfolding, the high energy peak is significantly suppressed.

\begin{figure}[!htb]
    \centering
    \begin{subfigure}{a}
    \includegraphics[width=0.45\textwidth]{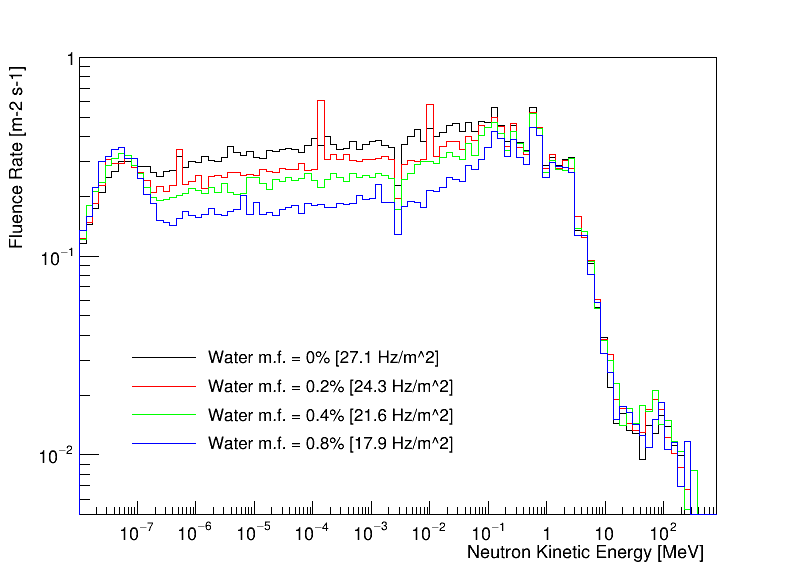}
    \end{subfigure}
    \begin{subfigure}{b}
    \includegraphics[width=0.45\textwidth]{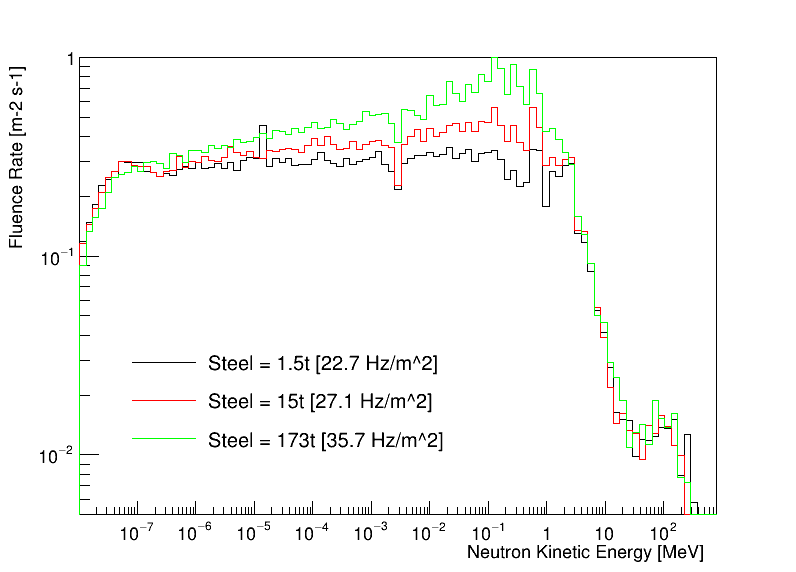}
    \end{subfigure}
    \caption{The Geant4 simulated effect of the variation of the water mass fraction in the concrete (a) and the mass of steel in the basement (b) on the neutron fluence rate and energy spectrum.}
    \label{fig:variation}
\end{figure}

There are several simplifications and assumptions in the configuration of the above Geant4 simulation, which may affect the reliability of the result. They are discussed in the following:
\begin{itemize}
    \item From the comparison of the thermal peak region between the unfolded and simulated spectra, it is inferred that the neutron modulation effect is much stronger in data than in simulation. The Hydrogen element is the best modulator. Thus it is tried to vary the assumed water mass fraction in the concrete up and down by a factor of 2 as well as to remove the water completely, which is shown in Figure\,\ref{fig:variation} a. As more water is added, the thermal peak gets more significant. Thus the underestimated thermal peak in the nominal Geant4 simulation implies that the water content should be increased. However, it shall have little impact on the unfolded spectrum since this misinformation will be corrected as the initial spectrum is iteratively update by data. Besides, the overall fluence rate is found to decrease by around 1 $Hz/m^2$ per 0.1\% of water added.
    \item The total mass of the steel in the basement is assumed to be 1.5 tons, which can be underestimated. Increasing this mass, the spectrum is found to be similar, except that the slope of the epithermal region is not so flat, as shown in Figure\,\ref{fig:variation} b. It can be easily understood: the steel is in the basement, thus neutrons generated from it do not have to travel through the concrete and are less moderated. As for the overall fluence rate, it is found to increase by 0.3 to 0.4 $Hz/m^2$ per ton of steel added.
    \item The dimension of the basement is assumed to be 5x5x6\,$m^3$, but in reality, it's much larger in the horizontal plane. Three additional simulations are performed with the basement dimension enlarged to 7x7x6, 9x9x6, and 11x11x6\,$m^3$, respectively. The resulting fluence rates and energy spectra are similar.
    \item Vacuum is assumed in the basement. If replacing vacuum with an air of 70\% humidity, the fluence rate and energy spectrum barely change.
    \item The same neutron may be counted more than once if it passes several virtual detectors. It is tried to change the size of the plane detector from 1x1\,$m^2$ to 0.7x0.7\,$m^2$. With such a smaller sensitive area, the multiple crossing probability of a neutron over the 5x5x5 detector array is significantly reduced, while the results stay the same.
    \item The overburden thickness is assumed to be 3.5 meters in the simulation. However, varying it by $\pm$ 0.5 meters barely changes the results.
\end{itemize}

\section{Summary}
\label{sec:1:sum}

Underground ambient neutrons are measured in terms of its fluence rate and kinetic energy spectrum in the basement where the TAO experiment will be located with an overburden of roughly 4 meters.
A Bonner sphere spectrometer is used and its count rates are unfolded with the iterative Maximum-Likelihood Expectation–Maximization method. Geant4 is used to simulate the cosmic ray muon-induced neutrons and provide the initial energy spectrum to the unfolding as well as to understand the unfolded results. 

The Geant4 simulation is introduced in details and varied systematically. The thermal neutron peak is sensitive to the water content in the concrete, while the epithermal region can be significantly affected if the mass of steel (or similar material which is a good neutron generation target) in the basement is varied. The variation of the dimension of the basement and the thickness of the overburden in a moderate range shows little impact on the simulated result.

The energy-integrated neutron fluence rate is measured to be 36.1 $\pm$ 4.7 $Hz/m^2$, which is three sigmas higher than the Geant4 simulation. The uncertainty is dominated by the overall scaling uncertainty of the Bonner sphere response functions and the Poisson statistical uncertainty of the raw counts. 
The integrated fluence rates from 10$^{-8}$\,MeV to 20\,MeV are compatible between data and simulation, which implies that the main discrepancy comes from the high-energy neutron region beyond 20\,MeV.
The possible sources of this discrepancy are discussed but no conclusion can be derived. 
The relative strength of the fluence rates between the thermal and epithermal regions is also mis-modeled, but can be explained by the inaccurate water content information in the concrete implemented in the simulation.
The unfolded energy spectrum is provided and shows similar deviations when compared with the Geant4 simulation.

\section*{Acknowledgments}
\label{sec:1:acknow}
This work was supported by the National Natural Science Foundation of China No. 11875282 and 12022505, the Strategic Priority Research Program of the Chinese Academy of Sciences, Grant No. XDA10011200, the CAS Center for Excellence in Particle Physics and the Youth Innovation Promotion Association of CAS.

\printcredits

\bibliographystyle{unsrtnat}

\bibliography{cas-refs}


\bio{}
Author biography without author photo.
Author biography. Author biography. Author biography.
\endbio

\bio{figs/pic1}
Author biography with author photo.
Author biography. Author biography. Author biography.
\endbio

\bio{figs/pic1}
Author biography with author photo.
Author biography. Author biography. Author biography.
\endbio

\end{document}

%% file: tabs/fluxes.tex
\begin{table}[htb]
\centering
\caption{The unfolded energy-integrated neutron fluence rates in the thermal peak, epithermal, fast peak, and high energy peak regions, as well as in the full energy range. The Poisson statistical and  systematic uncertainties are shown with the percentages as the corresponding relative uncertainties. The differences between the simulated and unfolded rates are shown and quantified in unit of standard deviation.}
\begin{tabular}{l|c|c|c|c|c|c}
   \hline
   \hline
   \multicolumn{2}{c|}{Energy Range (MeV)} & [1e-8,1e-6] & [1e-6,1e-1] & [0.1,20] & [20,800] & All \\ 
   \hline
   \multicolumn{2}{c|}{Unfolded Fluence Rate ($Hz/m^2$)} & 9.0& 8.2& 6.0& 12.8& 36.1\\ 
   \hline
   \multirow{6}{*}{Uncertainty} & Poisson & 1.2 (14\%)& 0.3 (3\%)& 0.6 (10\%)& 2.5 (19\%)& 2.8 (8\%)\\ 
   \cline{2-7}
 & Iteration Times & 0.2 (2\%)& 0.2 (3\%)& 0.7 (11\%)& 2.0 (15\%)& 1.4 (4\%)\\ 
   \cline{2-7}
 & Initial Spectrum & 1.8 (20\%)& 1.3 (16\%)& 0.7 (11\%)& 1.2 (9\%)& 0.0 (0\%)\\ 
   \cline{2-7}
  & BSS Scaling Factor & 0.9 (10\%)& 0.8 (10\%)& 0.6 (10\%)& 1.3 (10\%)& 3.6 (10\%)\\ 
   \cline{2-7}
   & All Systematics  & 2.0 (22\%)& 1.5 (19\%)& 1.1 (19\%)& 2.6 (20\%)& 3.9 (11\%)\\ 
   \cline{2-7}
   & Total & 2.4 (26\%)& 1.6 (19\%)& 1.3 (21\%)& 3.6 (28\%)& 4.8 (13\%)\\ 
   \hline
   \hline
   \multicolumn{2}{c|}{Simulated Fluence Rate ($Hz/m^2$)} & 4.6& 12.1& 4.8& 0.2& 21.6\\ 
   \hline
   \multicolumn{2}{c|}{Deviation wrt Unfolded} & -4.5 & 3.8 & -1.2 & -12.7 & -14.5 \\ 
   \hline
   \multicolumn{2}{c|}{Significance ($\sigma$)} & -1.9& 2.5& -1.0& -3.5& -3.0\\ 
   \hline
   \hline
\end{tabular}
\label{tab:fluxes}
\end{table}